# IS ESA RELATIONSHIP THE TOOL IN SEARCHING FOR INTERSTELLAR HETEROCYCLES?

E. E. Etim*[1], Inyang, E. J.[2], Ushie, O. A[1]., Mbakara, I. E.[3], Andrew, C., U. Lawal[1]

[1]Department of Chemical Sciences, Federal University Wukari, Nigeria
[2]Department of Pure and Industrial Chemistry University Of Nigeria, Nsukka
[3]Department of Chemistry, University of Ibadan, Ibadan

*Email: emmaetim@gmail.com

**Abstract:** Because of their importance in biological systems, in our understanding of the solar system and in other applications, seven heterocycles; furan, imidazole, pyridine, pyrimidine, pyrrole, quinoline and isoquinoline have been astronomically searched for in different molecular clouds with only the upper limits in the range of $4*10^{12}$ to $2.8*10^{21} cm^{-2}$ determined for their column densities in all the cases without any successful detection. Bothered by their unsuccessful detection, the energy, stability and abundance (ESA) relationship existing among interstellar molecules has been applied to examine if the these heterocycles were the best candidates for astronomical searches in terms of interstellar abundance in relation to other isomers of each group. High level quantum chemical calculations have been used to determine accurate enthalpies of formation for 67 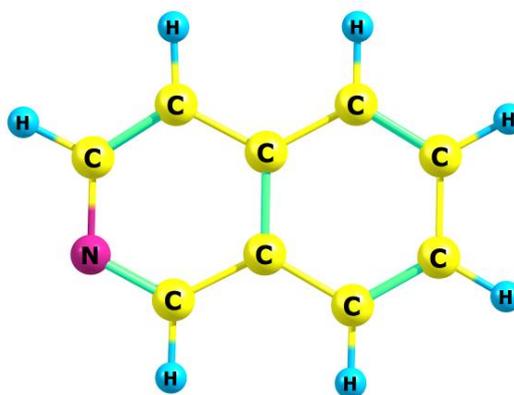 molecules from different isomeric groups of these heterocycles. From the results, all the 7 heterocycles so far searched are the best candidates for astronomically observations as they are molecules with the least enthalpies of formation in their respective groups and by extension, the most stable and probably the most abundant species in ISM. They remain the best possible molecules among other heterocycles to be observed soon with improved sensitivity of astronomical instruments, precise rest frequencies and proper choice of astronomical sources.

***Key words*:** ISM: abundance- ISM: atoms- ISM: molecules: physical data and processes: astrochemistry

**1. Introduction:** Despite the broad overlap of the different branches in chemical sciences which has hampered the clear division of these branches, heterocycles have emerged as a unique division in organic chemistry with vast applications and importance in diverse areas. Simply put, they are compounds containing at least one atom of carbon and at least one element other than carbon (which could be oxygen, sulphur or nitrogen), within a ring structure. Generally, organic ring compounds are known to play fundamental roles in terrestrial biochemistry and are believed to have been important ingredients in the earth's prebiotic chemistry. From the astrobiology and prebiotic chemistry perspectives, heterocycles are fundamental building blocks of biological systems as formation subunits of deoxyribonucleic acid (DNA) and ribonucleic acid (RNA). Medically, they are constituents of most drugs currently used as anti-HIV, antifungal, antimicrobial, antimalarial, antdiabetic, herbicidal, fungicidal, antibiotic, antidepressant, and antitumor agents (Dua et al. 2011; Valverde & Torroba 2005). Industrially, they are used as dyes, brightening, information storage and analytical agents. Heterocycles are also useful in other research areas such as polymer chemistry (as conjugated polymers), inorganic and supra molecular chemistry (as coordination compounds).

The presence of these molecules in carbonaceous chondrites suggests that molecules of biological interest could be formed in non-terrestrial environment through abiotic pathway. It is well known that many of the molecules found in meteorites originated in the interstellar or circumstellar medium. That the heterocycles have been detected in the meteorites is a further confirmatory test of their presence in ISM. Polymerization of small molecules such as ethyne ($C_2H_2$) and the incorporation of N-atoms via the substitution of ethyne by cyanic acid is considered as a possible formation route for the N-heterocycles. Both ethyne and cyanic acid are well known interstellar molecules with good abundance; hence the above formation route is very plausible (Ricca et al. 2001; Nuevo et al. 2012; Botta & Bada 2012; Snyder & Buhl 1971; Ridgway et al. 1976; Lacy et al. 1989).

Molecular formation processes in ISM are largely thermodynamically controlled, kinetics plays a role in very few cases; this fact is well shown to be true by the Energy, Stability and Abundance (ESA) Relationship existing among interstellar molecules. According to the ESA relationship, "*Interstellar abundances of related (isomers) species are directly proportional to their stabilities in the absence of the effect of interstellar hydrogen bonding*". (Etim & Arunan 2016). Isomerism is playing a vital role interstellar chemistry; astronomical observation of an isomer is always a pointer to the presence of other isomers though not yet detected, since isomers are believed to have a common precursor for their formation (Hollis 2005). The concept of isomerism can therefore be used to demystify some astronomical issues.

Seven heterocycles furan, imidazole, pyridine, pyrimidine, pyrrole, quinoline and isoquinoline are currently listed as non-detected interstellar molecules (http://www.astrochymist.org/astrochymist_nondetections.html) as a result of the unsuccessful astronomical searches for these molecules in different astronomical sources. The searches only yielded upper limits in the range of $4*10^{12}$ to $2.8*10^{21} cm^{-2}$ determined for their column densities in all the cases without any successful detection. This has resulted in questions like; were these heterocycles the best candidates for astronomical searches? Are there isomers that would have been better options for astronomical searches? In other to address these questions, we have considered 67 isomers from the six isomeric groups where these 7 heterocycles belong. High quantum chemical calculations have been applied to determine accurate enthalpies of formation for each of these isomers.

## 2 Computational details

The quantum chemical calculations reported in this work were carried out using the Gaussian 09 suite of programs (Frisch et al. 2009). By definition, the standard enthalpy of formation ($\Delta_f H^0$) of any molecule is the enthalpy change of the reaction by which it is formed from its constituent's elements. Among the different composite quantum chemical methods that are now used to accurately predict thermochemistry data, the G4 method has been found to be found to be very effective in predicting enthalpy of formation values to chemical accuracy in many molecules as reported in literatures ( Curtiss et al. 2007a,b; Etim and Arunan 2016). The reported values of the enthalpies of formation of all the molecules considered in this study were calculated from their atomization energies from the same level of theory (G4). For this calculation, the zero-point corrected standard enthalpies of formation of all the molecules considered in this study were calculated from the optimized geometries of the molecules at the level of theory mentioned above. The structures were found to be stationary with no imaginary frequency through harmonic vibrational frequency calculations.

### 2.1 Atomization energies and enthalpy of formation

The atomization energies represented as $D_o$ (since it is sometimes synonymously referred to as the total dissociation energies) were evaluated using the calculated values of energies (sum of electronic and zero-point energy corrections) with the methods described in the computational methods above. For a reaction,

2A + B → CD          (1)

The expression for computing the atomization energy of the molecule (CD) is given as;

$$\sum D_0(CD) = 2E_0(A) + E_0(B) - E_0(CD) \quad (2)$$

In calculating the enthalpy of formation ($\Delta_f H^0$) at 0 K for all the molecules reported in this study, the experimental values of standard enthalpy of formation of elements C, H, O, and N reported in literature (Curtiss et al. 1997) were used.

The enthalpy of formation at 0 K is calculated using the following expression:

$$\Delta H_f^0(CD, 0K) = 2\Delta H_f^0(A, 0K) + \Delta H_f^0(B, 0K) - \sum D_0(CD) \quad (3)$$

The enthalpy of formation at 298 K is calculated using the following expression:

$$\Delta H_f^0(CD, 298K) = \Delta H_f^0(CD, 0K) + (H_{CD}^O(298K) - H_{CD}^0(0K)) - \left[2\{H_A^0(298K) - H_A^0(0K)\} + \{H_B^0(298K) - H_B^0(0K)\}\right] \quad (4)$$

Table 1. Experimental $\Delta_f H^0$ (0K), of elements and $H^0$ (298K) – $H^0$ (0K)

From equation 4, $H_{CD}^{O}(298K) - H_{CD}^{0}(0K)$ is defined as $H_{corr}$ - $E_{zpe}$. Where $H_{corr} = E_{tot} + k_B T$ and $E_{tot} = E_t + E_r + E_v + E_e$ (Etim & Arunan 2015).

## 3. Results and Discussion

The reported enthalpies of formation in this study are in kcal/mol and at 298.15K. The results obtained from the 6 different isomeric groups where these 7 heterocycles belong are presented (in tables and figures) and discussed accordingly.

**3.1 $C_3H_4N_2$ Isomers:** Table 1 presents the optimized structures and enthalpies of formation for the 13 isomers of the $C_3H_4N_2$ isomeric group in decreasing order of enthalpy of formation. This isomeric group consists of important biological molecules; the imidazole ring structure occurs in fundamental biological molecules such as histidine (an amino acid) and purines; imidazolium and pyrazolium-based ionic surface are now receiving attention of researches as good materials for biosensing; pyrazole and its derivatives are important constituents of many drugs because of their muscle relaxing, antidiabetic, analgesic, anti-inflammatory, and other biological activities (Irvine et al. 1981; Noe & Fowden 1959; Ratel et al. 2013).

As important as the $C_3H_4N_2$ isomers are, only imidazole has been astronomically searched for (Irvine et al. 1981; Dezafra et al. 1971). In accordance with the ESA relationship, among the isomers of $C_3H_4N_2$ isomeric group, imidazole has least enthalpy of formation (Table 1) as compared to other isomers of the group and as such it is expected to be the most stable isomer of the group and probably the most abundant isomer of the $C_3H_4N_2$ group in the interstellar medium (ISM). The astronomical search of imizaloe in the Sgr B2 molecular cloud only lead to an upper limit of the order of $10^{15}$ cm$^{-2}$ determined for its column density in Sgr B2 (Dezafra et al. 1971). Irvine et al. (1981) reported the upper limit in the range of 0.6 to 100 * $10^{13}$ cm$^{-2}$ for the column density of imidazole in different astronomical sources. Both searches could not yield any successful detection of imidazole in ISM. In comparison to other isomers of the group, imidazole remains the most potent candidates for astronomical observation being the most stable and probably the most abundant isomer of the group in ISM. With the range of upper limit so far determined for its column density, its astronomical observation will require highly sensitive astronomical instruments, more precise transition frequencies and appropriate choice of astronomical sources.

Figure 1 shows the plot of the enthalpy of formation for the various isomers of $C_3H_4N_2$ with the astronomically searched isomer indicated with green while the non-astronomically searched isomers are indicated with red. It is crystal clear from the plot that the only astronomically searched isomer is also the isomer with the least enthalpy of formation, making it the best choice for astronomical observation. This shows the direct application of the ESA relationship as a tool in searching for interstellar molecules.

Table 1: Structure and $\Delta_f H^O$ for $C_3H_4N_2$ isomers

| Molecule | Structure | Enthalpy of formation (kcal/mol) |
|---|---|---|
| Diazocyclopropane | 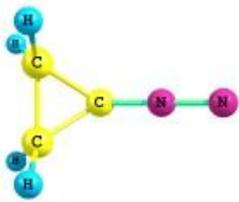 | 91.710 |
| Pyrazolium | 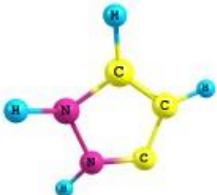 | 86.986 |
| 1-Azaridinecarbonitrile | 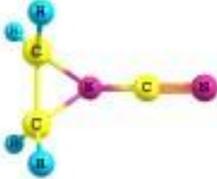 | 65.786 |
| 3H-Pyrazole | 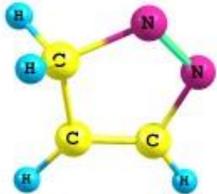 | 63.298 |
| 2-Azaridinecarbonitrile | 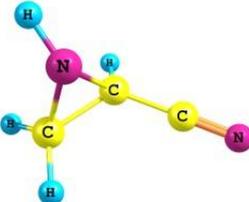 | 62.042 |
| 4H-Pyrazole |  | 59.291 |

|  | 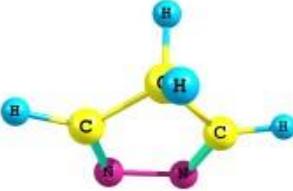 |  |
| --- | --- | --- |
| Imidazolium | 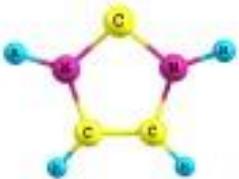 | 53.781 |
| Methylamino acetonitrile | 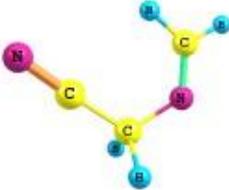 | 50.649 |
| 3-iminopropanenitrile | 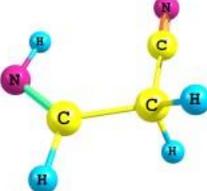 | 44.297 |
| 2-aminoacrylonitrile | 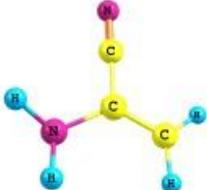 | 43.493 |
| Pyrazole |  | 36.958 |

|  | 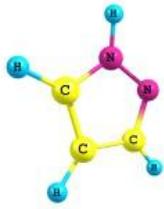 |  |
|---|---|---|
| 3-aminoacrylonitrile | 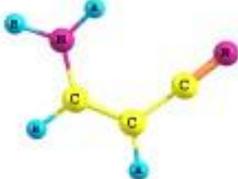 | 35.805 |
| **Imidazole** | 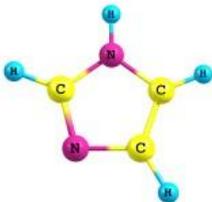 | **26.641** |

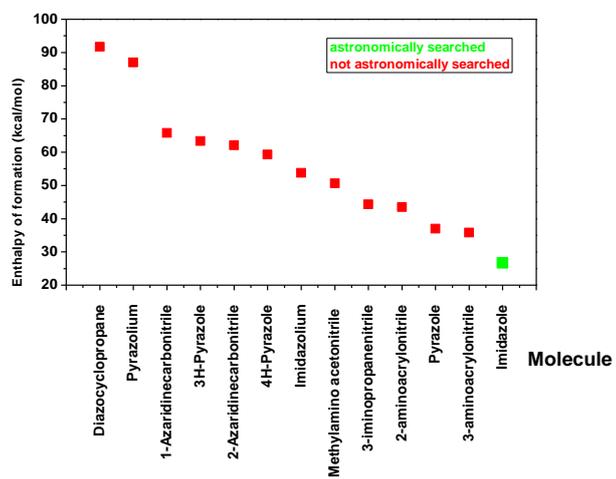

Figure 1: Plot showing the $\Delta_f H^O$ for $C_3H_4N_2$ isomers

**3.2 C$_5$H$_5$N Isomers:** Azafulvene is an inseparable partner in pyrrole chemistry, having been postulated as an intermediate in the nucleophilic substitution reaction at the 2-pyrrolylmethyl position. The structure of 1-cyanobicyclo[1.1.0]butane makes it a choice monomer in polymer chemistry. Pyridine has a long-familiar history of applications in the biological and chemical systems (Setsune et al. 2006; Linnell 1960; Pearson & Williams 1953). Broadly speaking, all the isomers of the C$_5$H$_5$N (Table 2 and Figure 2) group have important applications in diverse areas. However, when it comes to astronomical observations, other parameters are very crucial. These include energy, stability and interstellar abundance which are the components of the ESA relationship. Three different sets of scientists have reported unsuccessful searches for pyridine from different astronomical sources (Batchelor et al. 1973; Simon and Simon 1973; Charnley et al. 2005). Upper limits in the range of $7.3*10^{12}$ to $2.5*10^{15}$ cm$^{-2}$ have been determined for the column abundance of pyridine in the different astronomical sources searched.

The optimized structures and enthalpies of formation for the different isomers of C$_5$H$_5$N isomeric group are shown in Table while Figure 2 depicts the plot of the enthalpies of formation for these isomers. The wide difference between the enthalpy of formation of the most stable isomer; pyridine, which is also the only astronomically searched isomer (indicated with green) and the other isomers which have not been astronomically searched (indicated with red) is well pictured in Figure 2. In conformity with the ESA relationship, pyridine is the only isomer of the group that has been astronomically searched. Pyridine is undisputedly the best candidate for astronomical observation in comparison with other isomers of this group. It has the least enthalpy of formation; hence it is more stable and probably more abundant in ISM than other isomers of the group. The successful astronomical observation of pyridine is a function time with the current rapid development in astronomical and spectroscopic equipments that will culminate into better sensitivity of astronomical instruments and more accurate rest frequencies.

Table 2: Structure and $\Delta_f H^O$ for C$_5$H$_5$N isomers

| Molecule | Structure | Enthalpy of formation (kcal/mol) |
|---|---|---|
| 1-cyanobicyclo[1.1.0]butane | 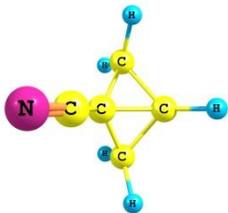 | 84.175 |
| 4-cyano-1-butyne | 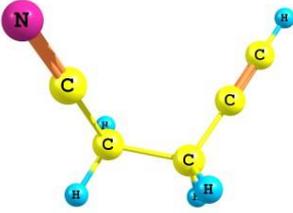 | 77.420 |
| 2,4-cyclopentadiene-1-imine | 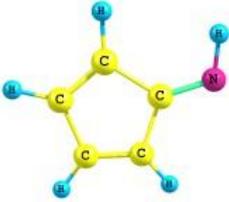 | 61.170 |
| 2-methylene-3-butenenitrile | 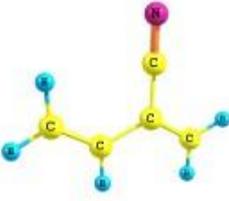 | 58.171 |
| Azafulvene | 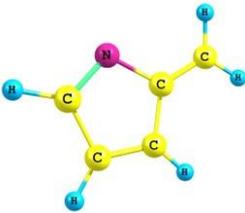 | 56.062 |
| 1-cyano-1,3-butadiene | 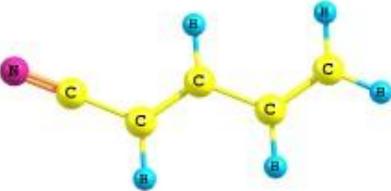 | 55.518 |
| *Pyridine* | | *28.737* |

|  | 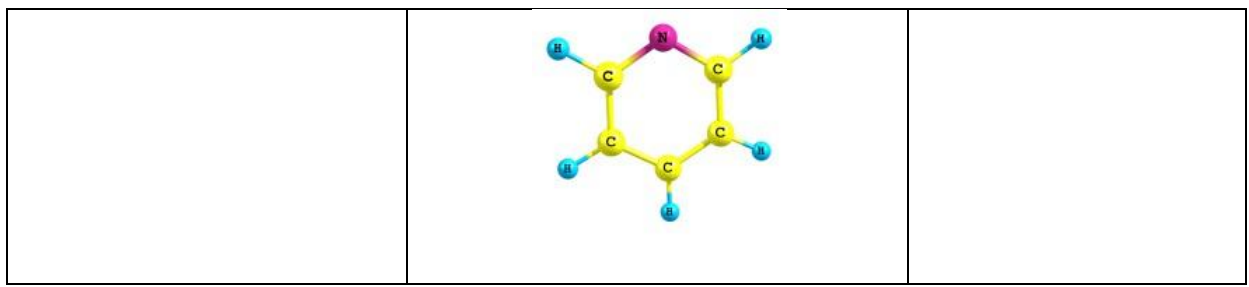 |  |

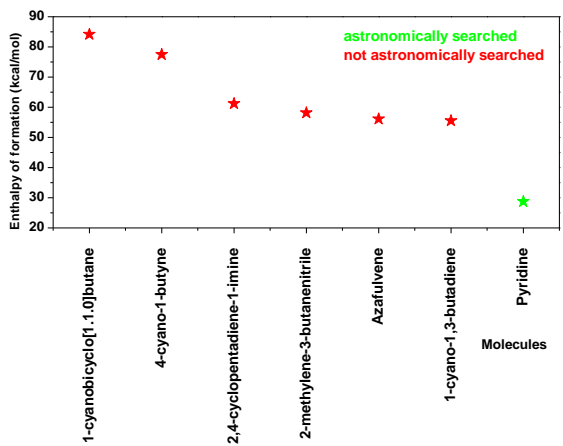

Figure 2: Plot showing the $\Delta_f H^O$ for $C_5H_5N$ isomers

3.3 **$C_4H_4N_2$ Isomers**: The enthalpy of formation calculated with the G4 method employed in this study for the different isomers of the $C_4H_4H_2$ isomeric group and their optimized structures are presented in table 3. Despite the importance of all the isomers of this group, only pyrimidine has been astronomically searched for in different molecular clouds. The choice for the astronomical search of pyrimidine at the expense of other is easily understood from the ESA relationship. Pyrimidine has the least enthalpy of formation as compared to other isomers of the group and as such it is expected to be more stable and probably more abundant in the interstellar space than other isomers of the group. Figure 3 nicely displays the enthalpy of formation of these isomers with the astronomically searched isomer which is also the isomer with the least enthalpy of formation indicated with green while others are indicated with red.

The astronomical searches for pyrimidine in different molecular clouds yielded upper limits of $1.7*10^{14}$ cm$^{-2}$ for Sgr B2(N), $2.4*10^{14}$ cm$^{-2}$ for Orion KL and $3.4*10^{14}$ cm$^{-2}$ for W51 e1/e2 as column abundance determined for pyrimidine in the respective molecular clouds (Kuan et al. 2003), without any successful detection. Among all the isomers of this group, pyrimidine remains the most possible molecule for astronomical observation in the nearest future.

Table 3: Structure and $\Delta_f H^O$ for $C_4H_4N_2$ isomers

| Molecule | Structure | Enthalpy of formation (kcal/mol) |
|---|---|---|
| 1,2-diisocyanoethane | 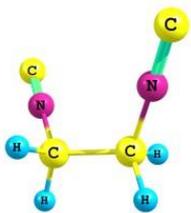 | 92.445 |
| 1,3-butadiene-1,4-diimine | 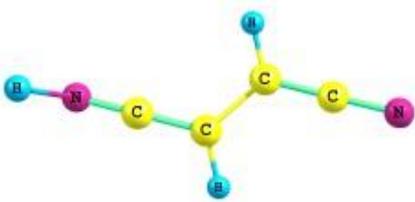 | 88.011 |
| 4-amino-2-butynenitrile | 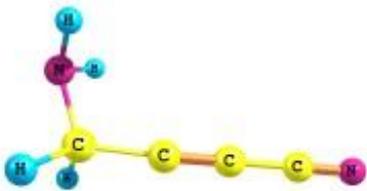 | 83.014 |
| Iminopyrrole | 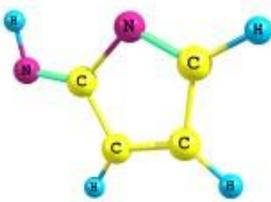 | 62.267 |
| 2-methylene-2H-imidazole | 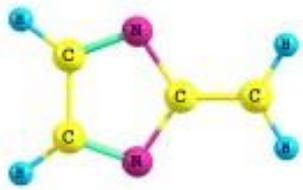 | 60.147 |
| Pyridazine | | 59.052 |
| 12 | 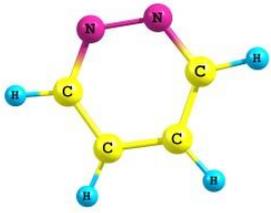 | |
| 1,1-dicyanoethane | 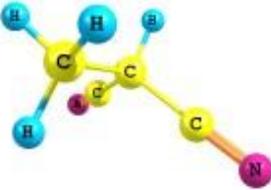 | 54.713 |
| Pyrazine | 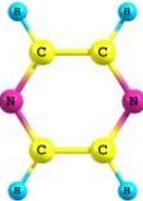 | 41.006 |
| *Pyrimidine* | 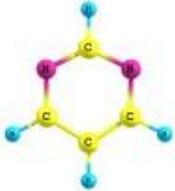 | **37.114** |

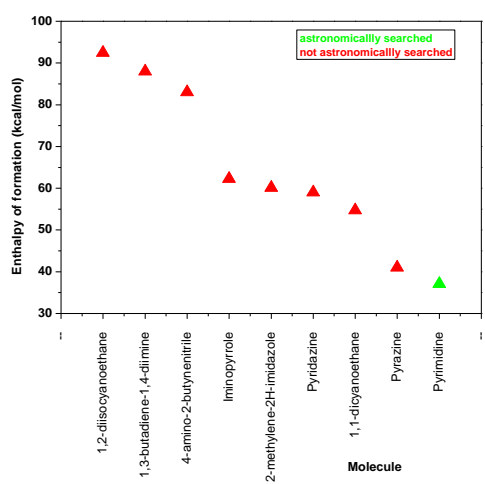

Figure 3: Plot showing the $\Delta_f H^O$ for $C_4H_4N_2$ isomers

3.4 **$C_4H_5N$ isomers:** Eleven isomers of the $C_4H_5N$ isomeric group with their corresponding optimized nstructures and enthalpies of formation calculated with the G4 method are shown in Table 4. These isomers are widely studied in different research areas because of their various applications. Most of them are extensively used in organic synthesis, e.g, pyrrole, ally cyanide etc. Ally cyanide is a good cross-linking agent in some polymers. It has also been proposed as an additive in propylene carbonate-based electrolytes for graphite anode to prevent exfoliation of the anode via film-forming. In biological systems, ally cyanide has been shown to induce antioxidant and detoxification enzymes (Segall & Zare 1988; Deniau et al. 1992; Tanii et al. 2004, 2008). As much as these isomers are important in various aspects, when it comes to astronomical searches, other parameters emerge as the priority. These are the energy, stability and interstellar abundance as described in the ESA relationship.

Just as in the previous cases discussed above, the only astronomically searched isomer of this group; pyrrole is also the isomer with the least enthalpy of formation as compared to other isomers of the group. Astronomical searches for pyrrole have been reported by two groups of scientists (Kutner et al. 1980; Myers et al. 1980), with the upper limits in the range of 3 to 10 $*10^{13}$ cm$^{-2}$ for Sgr B2 and $4*10^{12}$ cm$^{-2}$ for TMC determined for the column density of pyrrole in the respective astronomical sources without any successful detection of pyrrole. In the plot of the enthalpy of formation of these isomers (Figure 4), the astronomically searched isomer; pyrrole (indicated with green), is conspicuously different from the other isomers (indicated with red) in terms of the magnitude of the enthalpy of formation. Among all the isomers of this group, pyrrole still remains as the most potent candidate for astronomical observation being the most stable and probably the most abundant isomer of the group in ISM.

Table 4: Structure and $\Delta_f H^O$ for $C_4H_5N$ isomers

| Molecule | Structure | Enthalpy of formation (kcal/mol) |
|---|---|---|

| 2-vinyl-2H-azirene | 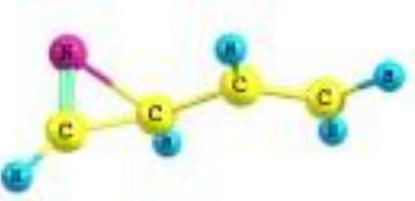 | 78.561 |
| --- | --- | --- |
| Isocyanocyclopropane | 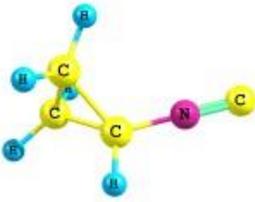 | 65.289 |
| Ally isocyanide | 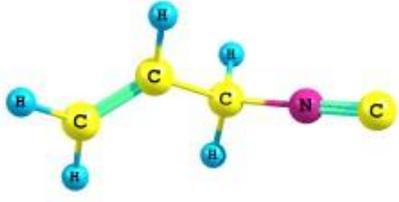 | 61.986 |
| N-vinylethyleneimine | 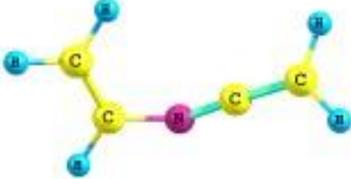 | 60.971 |
| Cyanocyclopropane | 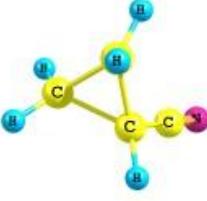 | 43.605 |
| 2H-pyrrole | 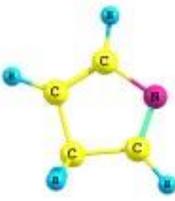 | 40.051 |
| 3H-pyrrole | | 40.049 |
| 15 | 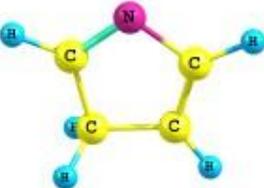 | |
| Ally cyanide | 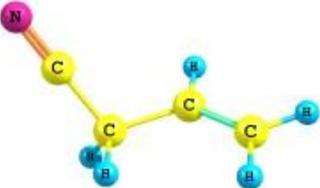 | 39.742 |
| 2-cyanopropene | 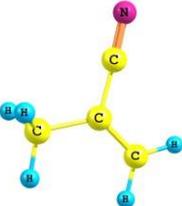 | 34.031 |
| 2-butenenitrile | 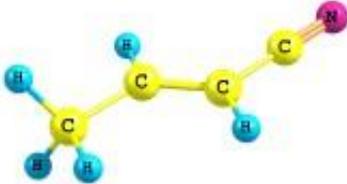 | 32.787 |
| **Pyrrole** | 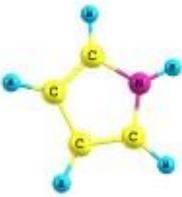 | **24.094** |

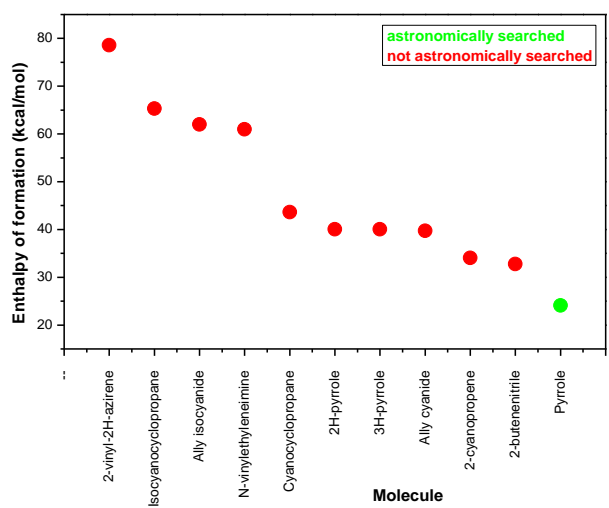

Figure 4: Plot showing the $\Delta_f H^O$ for $C_4H_5N$ isomers

**3.5 $C_9H_7N$ Isomers**: Unlike in the previous cases where only one isomer has been astronomically searched for from each isomeric group, in the $C_9H_7N$ isomeric group, 2 isomers have been astronomically searched. *Will this group still follow the ESA relationship?* Table 5 gives the optimized structures and enthalpies of formation for the 13 isomers of the $C_9H_7N$ isomeric group. These isomers are well studied for their diverse applications in several fields.

With respect to astronomical searches, the trend is the same as those discussed above. Quinoline and its closest chemically related isomer, isoquinoline are the only two isomers of this group that have been astronomically searched. As listed in Table 5, these two isomers are also the isomers with the least enthalpies of formation in comparison with other isomers of the group and as such are expected to be the most stable and probably the most abundant isomers of the group in ISM. Charnley et al. (2005) reported unsuccessful astronomical searches for quinoline and isoquinoline in three different astronomical sources with upper limits in the range of $1.9*10^{13}$ to $1.2*10^{18}$ cm$^{-2}$ for quinoline and $2.9*10^{13}$ to $2.8*10^{21}$ cm$^{-2}$ for isoquinoline determined for their column densities in the different molecular clouds searched. Figure 5 shows the plot of the enthalpies of formation of these isomers.

Though yet to be successfully observed in ISM, quinoline and isoquinoline remain the most likely isomers for astronomical observation from this group of isomers. They will probably require more sensitive astronomical tools, accurate rest frequencies and excellent choice of astronomical sources for their observations.

Table 5: Structure and $\Delta_f H^O$ for $C_9H_7N$ isomers

| Molecule | Structure | Enthalpy of formation (kcal/mol) |
|---|---|---|
| 3,4-diethynyl-1-methyl-1H-pyrrole | 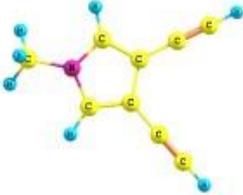 | 136.304 |
| 2,5-diethynyl-1-methyl-1H-pyrrole | 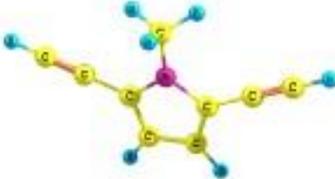 | 133.40 |
| 3-phenyl-2-propyn-1-imine | 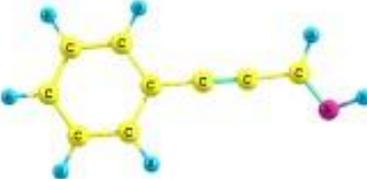 | 98.259 |
| 1-isocyano-2-vinyl-benezen | 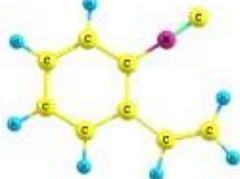 | 89.268 |
| 1-isocyano-4-vinyl-benzene | 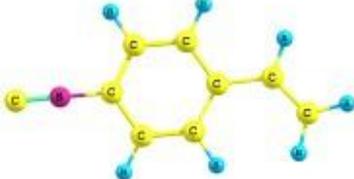 | 87.015 |
| Bicyclo[4.2.0]1,3,5-triene-7-carbonitrile | 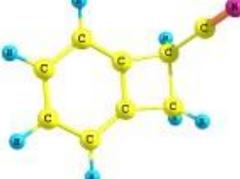 | 82.259 |
| Atroponitrile | 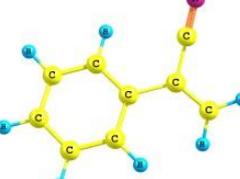 | 68.951 |
| 3-vinylbenzonitrile | 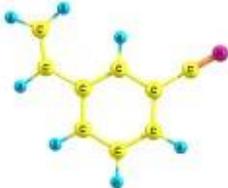 | 67.432 |
| 4-vinylbenzonitrile | 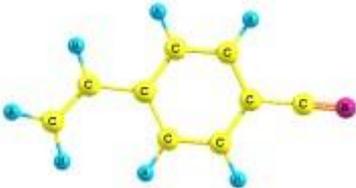 | 66.918 |
| 3-methylene-3H-indole | 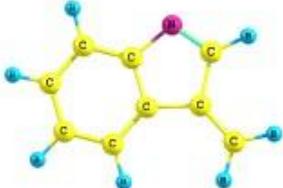 | 66.643 |
| Cyanostyrene | 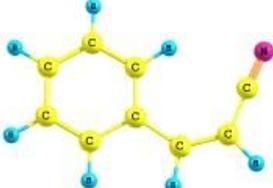 | 66.518 |
| **Isoquinoline** | 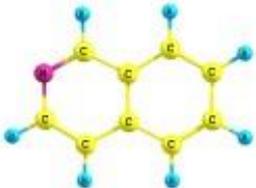 | **45.956** |
| **Quinoline** | 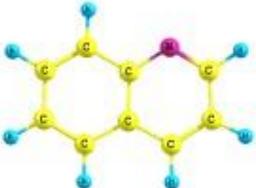 | **44.781** |

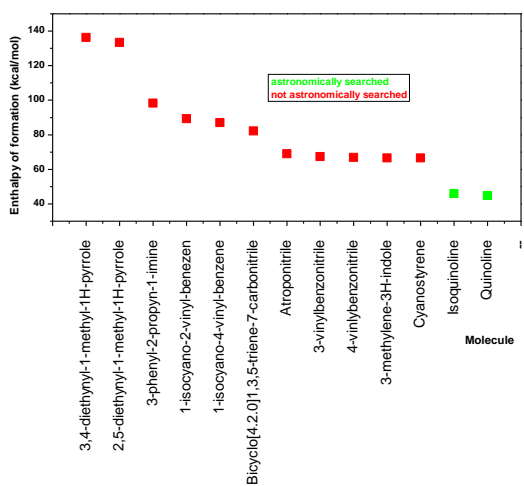

Figure 5: Plot showing the $\Delta_f H^O$ for $C_9H_7N$ Isomers

**3.6 $C_4H_4O$ Isomers:** All our discussions so far (from subsections 3.1to 3.5) have been on one type of heterocycles; the N-heterocycles. It will be good to see if the ESA relationship is also used as a tool in searching for other heterocycles not only N-heterocycles as seen thus far. The $C_4H_4O$ isomeric group contains oxygen rather than nitrogen as in the previous groups. The enthalpies of formation for 14 isomers of this group calculated with the G4 method and the optimized structures of these isomers are presented in Table 6 below.

The trend is exactly the same as discussed for the N-heterocycles. Furan is the only astronomically searched isomer of this group, it is also the isomer with the least enthalpy of formation with marked difference in magnitude even with the next isomer; vinylketene. This is well pictured in Figure 6. Furan's stability and its expected interstellar abundance in comparison with other isomers of the group warranted its astronomical searches. Dezafra et al. (1971) and Kutner et al. (1980) have reported unsuccessful searches for furan in different astronomical sources. The upper limits determined for the column density of furan in Orion A and Sgb B2 are $7*10^{13}$ cm$^{-2}$ and $2*10^{16}$ cm$^{-2}$ respectively. Furan remains a sure candidate for astronomical observation as compared to every other isomer of the $C_4H_4O$ group. Figure 6 nicely depicts the plot of the standard enthalpies of formation for the C4H4O isomers with the astronomically searched isomer, furan indicated in green while others are shown in red.

Table 6: Structure and $\Delta_f H^O$ for $C_4H_4O$ Isomers

| Molecule | Structure | Enthalpy of formation (kcal/mol) |
|---|---|---|

| Ethynyloxy ethene | 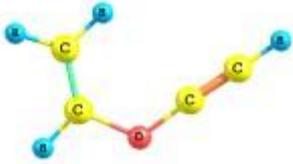 | 46.776 |
| Cyclopropene-1-carbaldehyde | 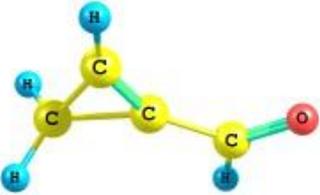 | 39.853 |
| 2-Cyclopropene-1-carbaldehyde | 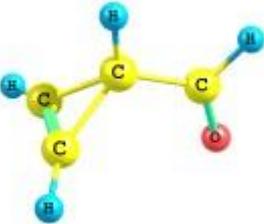 | 38.781 |
| 1-buten-3-yn-2-ol | 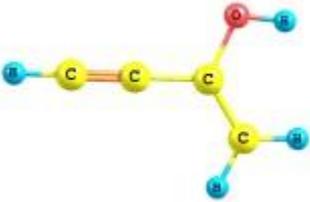 | 30.702 |
| But-3-ynal | 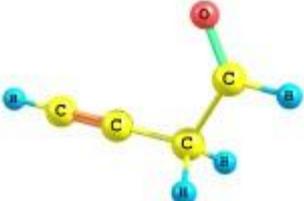 | 24.553 |
| 2-methyl-2-cyclopropenemethanone | 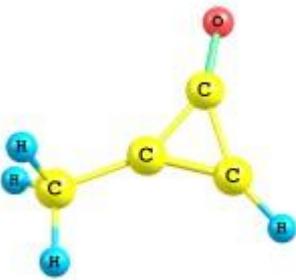 | 21.003 |
| Cyclopropylienemethanone | 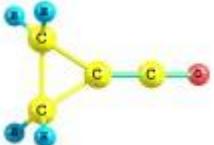 | 19.320 |
| 3-butyn-2-one | 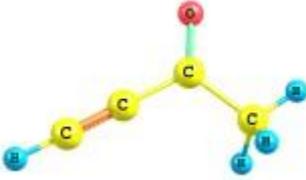 | 18.218 |
| 2-butynal | 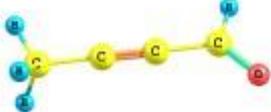 | 17.161 |
| 1,2-butadienal | 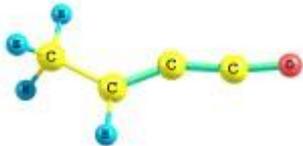 | 13.283 |
| 2,3-butadienal | 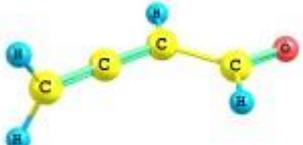 | 13.116 |
| 2-cyclobutene-1-one | 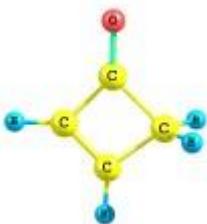 | 9.180 |
| Vinylketene | 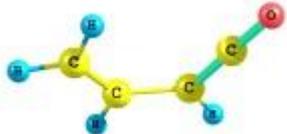 | 3.563 |
| Furan | 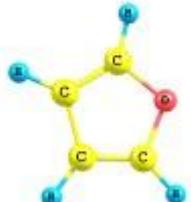 | -9.261 |

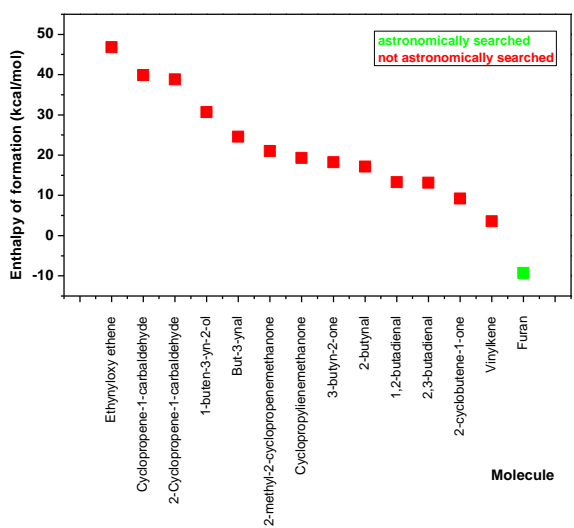

Figure 6: Plot showing the $\Delta_f H^O$ for $C_4H_4O$ Isomers

**Conclusion:** All the heterocyclic molecules that have so far been searched for in different astronomical sources have been examined in relation to their isomers using high level chemical quantum calculation. There has not been any successful astronomical observation of any of these molecules; they remain the best candidates for astronomical observation in their respective groups. This is because; all the heterocycles so far searched for are the ones with the least enthalpy of formation in their respective isomeric groups and as such are the most stable and probably the most abundant in ISM. Their delayed successful astronomical observation could be due to several factors; low resistant to photodissociation, large partition function, low abundance in ISM, etc. The astronomical observation of these molecules will demand more sensitive astronomical equipments, precise transition frequencies, proper choice of astronomical sources, among others. The consistency of the choice of these heterocycles for astronomical searches in each of the groups in accordance with the ESA relationship has led to the question; I*s ESA relationship the tool in searching for interstellar heterocycles?*

**Acknowledgement:** EEE acknowledges a research fellowship from the Indian Institute of Science, Bangalore.